%% file: main.tex
\documentclass[sigconf]{acmart}

\acmConference[SecureComm 2024]{20th EAI International Conference on Security and Privacy in Communication Networks}{October 2024}{Dubai, United Arab Emirates}

\setcopyright{none}
\renewcommand\footnotetextcopyrightpermission[1]{} 

\input{preamble}

\title{Is This the Same Code? A Comprehensive Study of Decompilation Techniques for WebAssembly Binaries}

\author{Wei-Cheng Wu}
\email{wei-cheng.wu.gr@darmouth.edu}
\affiliation{%
  \institution{Dartmouth College}
  \city{Hanover}
  \state{NH}
  \country{USA}
}

\author{Yutian Yan}
\email{yutianya@usc.edu}
\affiliation{%
  \institution{University of Southern California}
  \city{Los Angeles}
  \state{CA}
  \country{USA}
}

\author{Hallgrimur David Egilsson}
\email{egilsson@usc.edu}
\affiliation{%
  \institution{University of Southern California}
  \city{Los Angeles}
  \state{CA}
  \country{USA}
}

\author{David Park}
\email{dpark946@usc.edu}
\affiliation{%
  \institution{University of Southern California}
  \city{Los Angeles}
  \state{CA}
  \country{USA}
}

\author{Steven Chan}
\email{sychan@usc.edu}
\affiliation{%
  \institution{University of Southern California}
  \city{Los Angeles}
  \state{CA}
  \country{USA}
}

\author{Christophe Hauser}
\email{christophe.hauser@dartmouth.edu}
\affiliation{%
  \institution{Dartmouth College}
  \city{Hanover}
  \state{NH}
  \country{USA}
}

\author{Weihang Wang}
\email{weihangw@usc.edu}
\affiliation{%
  \institution{University of Southern California}
  \city{Los Angeles}
  \state{CA}
  \country{USA}
}

\begin{abstract}
WebAssembly (abbreviated WASM) is a low-level bytecode language designed for client-side execution in web browsers. 
As WASM continues to gain widespread adoption and its security concerns, the need for decompilation techniques that recover high-level source code from WASM binaries has grown. 
However, little research has been done to assess the quality of decompiled code from WASM.
This paper aims to fill this gap by conducting a comprehensive comparative analysis between decompiled C code from WASM binaries and state-of-the-art native binary decompilers. 
To achieve this goal, we presented a novel framework for empirically evaluating C-based decompilers from various aspects, thus assessing the proficiency of WASM decompilers in generating readable and correct code when compared to native binary decompilers.
Specifically, we evaluated the decompiled code's \textit{correctness}, \textit{readability}, and \textit{structural similarity} with the original code from current WASM decompilers.
We validated the proposed metrics' practicality in decompiler assessment 
and provided insightful observations regarding the characteristics and constraints of existing decompiled code.
By encouraging improvements in these tools, 
we seek to enhance their use in critical tasks such as auditing and sandboxing third-party libraries. 
This, in turn, contributes to bolstering the security and reliability of software systems that rely on WASM and native binaries.
\end{abstract}

\begin{document}

\ccsdesc[300]{Security and privacy~Software reverse engineering}
\keywords{
  WebAssembly, WASM, Decompiler, Reverse Engineering
}


\maketitle

\input{Intro.tex}

\input{Motivation}

\input{Detail.tex}

\input{Implementation}
\input{Evaluation}

\input{Discussion}

\input{Related}

\input{Conclusion}

\newpage
\bibliographystyle{ACM-Reference-Format}
\bibliography{main}

\end{document}

%% file: preamble.tex
\usepackage[utf8]{inputenc}
\usepackage{amsmath,amsfonts}
\usepackage{algorithmic}
\usepackage{algorithm}
\usepackage{array}
\usepackage[caption=false,font=normalsize,labelfont=sf,textfont=sf]{subfig}
\usepackage{textcomp}
\usepackage{stfloats}
\usepackage{url}
\usepackage{verbatim}
\usepackage{graphicx}

\usepackage{xcolor}
\usepackage{xspace}
\usepackage{svg}
\usepackage{pifont}  
\usepackage{booktabs} 
\usepackage{multirow}
\usepackage{wrapfig}
\usepackage{listings}
\usepackage{tikz}
\usepackage[framemethod=tikz]{mdframed}
\usepackage[most]{tcolorbox}
\usepackage{soul}
\usepackage{tabularx}
\usepackage{subcaption}
\usepackage[export]{adjustbox}

\usepackage{hyperref}   
\hypersetup{colorlinks = true, 
    urlcolor = black, 
    linkcolor = winered, 
    citecolor = winered 
}

\usepackage{enumitem}
\setlist[itemize]{align=parleft,left=0pt..1em}

\usepackage{framed}

\newcommand{\codeword}[1]{%
\texttt{\textcolor{black}{#1}}%
}

\setstcolor{red}

\usepackage{colortbl}

\definecolor{codegreen}{rgb}{0,0.6,0}
\definecolor{codepurple}{rgb}{0.58,0,0.82}
\definecolor{backcolour}{rgb}{0.95,0.95,0.92}
\definecolor{winered}{rgb}{0.7,0,0}
\definecolor{darkpastelgreen}{rgb}{0.01, 0.75, 0.24}
\definecolor{cadmiumgreen}{rgb}{0.0, 0.42, 0.24}
\definecolor{brickred}{rgb}{0.8, 0.25, 0.33}
\definecolor{cornellred}{rgb}{0.7, 0.11, 0.11}
\definecolor{burgundy}{rgb}{0.5, 0.0, 0.13}
\definecolor{frenchblue}{rgb}{0.0, 0.45, 0.73}
\definecolor{light-gray}{gray}{0.92}
\definecolor{lightlight-gray}{gray}{0.97}
\definecolor{codegray}{gray}{0.90}
\definecolor{inputgray}{gray}{0.90}
\definecolor{darkgreen}{RGB}{40,125,40}
\definecolor{amethyst}{rgb}{0.5, 0.3, 0.7}
\definecolor{dkgreen}{rgb}{0,0.6,0}
\definecolor{gray}{rgb}{0.5,0.5,0.5}
\definecolor{mauve}{rgb}{0.58,0,0.82}

\definecolor{shadecolor}{named}{light-gray}

\fontfamily{lmtt}\selectfont
\definecolor{dkgreen}{rgb}{0,0.6,0}
\definecolor{gray}{rgb}{0.5,0.5,0.5}
\definecolor{mauve}{rgb}{0.58,0,0.82}
\definecolor{am-green}{HTML}{1FCA23}
\definecolor{am-pink}{HTML}{FD42FC}
\definecolor{am-blue}{HTML}{1683FB}
\definecolor{cadmiumgreen}{rgb}{0.0, 0.42, 0.24}
\colorlet{mgray}{gray!90!black}

\lstdefinelanguage{JavaScript}{
  keywords={typeof, new, true, false, catch, function, return, null, catch, switch, var, const, let, if, in, while, do, else, case, break},
  keywordstyle=\color{blue}\bfseries,
  ndkeywords={class, export, boolean, throw, implements, import, this, self},
  ndkeywordstyle=\color{darkgray}\bfseries,
  identifierstyle=\color{black},
  sensitive=false,
  comment=[l]{//},
  morecomment=[s]{/*}{*/},
  commentstyle=\color{purple}\ttfamily,
  stringstyle=\color{red}\ttfamily,
  morestring=[b]',
  morestring=[b]"
}

\lstdefinelanguage{WAT}{
  keywords={module, type, func, param, result, call_indirect, call, table, export, import, elem},
  keywordstyle=\color{blue},
  identifierstyle=\color{black},
  sensitive=false,
  comment=[l]{;;},
  commentstyle=\color{purple}\ttfamily,
  stringstyle=\color{red}\ttfamily,
  morestring=[b]',
  morestring=[b]"
}

\global\mdfdefinestyle{rtboxstyle}{%
linecolor=black,%
leftmargin=0cm,rightmargin=0cm,linewidth=0.5pt,
roundcorner=3,
skipbelow=0pt,backgroundcolor=lightlight-gray
}

\def\BibTeX{{\rm B\kern-.05em{\sc i\kern-.025em b}\kern-.08em
    T\kern-.1667em\lower.7ex\hbox{E}\kern-.125emX}}

%% file: Intro.tex
\section{Introduction}
\label{sec:intro}

WebAssembly (WASM) is a portable, low-level language designed for near-native execution on the web~\cite{webassembly_introduction}. 
Since it was first introduced by Haas et al. in 2017~\cite{haas_bringing_2017} and initially developed for web browsers~\cite{webassembly_support}, 
its application has extended to diverse areas, 
including Internet of Things~\cite{warduino, aerogel_iot}, 
mobile devices~\cite{webassembly_mobile}, smart contracts~\cite{smart_contract}, 
and its dedicated runtime environments~\cite{wasmtime, wasmer}. 
Notably, WASM is commonly used as a compilation target for popular high-level languages like C, C++, and Rust~\cite{empirical}.

Despite its reputation for safety, WASM binaries still exhibit security vulnerabilities. 
Around 65\% of WASM binaries suffer from classic stack-based exploits commonly found in native binaries~\cite{empirical}, 
such as stack-based buffer overflows~\cite{ncc_group, forcepoint} and stack overflows~\cite{empirical}. 
The absence of standard memory protections, like stack canaries or guard pages, renders these vulnerabilities effective. 

Given the growing adoption of WASM, inspecting third-party binaries for potential security vulnerabilities has become imperative. 
However, the low-level nature of WASM bytecode makes it challenging to audit compared to high-level code, such as C. 
Additionally, 28.8\% of WASM binaries are minified \cite{empirical}, 
obfuscating variable/function names and making manual inspection cumbersome. 
To address these challenges, security experts can leverage decompilers to analyze high-level code instead of grappling with thousands of lines of minified low-level WASM code.

However, WASM decompilers have received less attention than decompilers designed for native binaries. 
Over the years, significant progress has been made in developing powerful native binary decompilers that can accurately generate decompiled code for C and C++ programs. 
Recent studies have also focused on enhancing the readability of decompiled code \cite{decompiler_high_readability, evolving_exact_decompilation}.

To this end, we perform a comprehensive study to assess the effectiveness of state-of-the-art WASM decompilers.

We will be approaching from two directions, as there are two different types of decompilers for WASM: 
decompilers tailored for readability, and decompilers focus on correctness, i.e., the decompiled code adheres to the behavior of the original WASM program.
Their performance is compared with off-the-shelf native binary decompilers \cite{ghidra, retdec}. 
To evaluate these decompilers, 
we utilize various widely-used complexity metrics for source code and adopt methodologies presented in previous studies \cite{schulte2018evolving, yakdan2015no, liu2020far}. 
Our study focuses on the following three aspects: 
\begin{itemize}
  \item \textbf{Correctness} of the decompiled code (Section~\ref{subsec:detail_correctness});
  \item \textbf{Readability} of the decompiled code (Section~\ref{subsec:detail_readability});
  \item \textbf{Structural similarity} between the decompiled code and the original code (Section~\ref{subsec:detail_similarity}).
\end{itemize}
 
With our research, we aim to draw attention to the capabilities of WASM decompilers and the performance of native binary decompilers. 
By encouraging improvements in these tools, 
we seek to enhance their use in essential tasks such as auditing and sandboxing third-party libraries \cite{rlbox}. 
This, in turn, contributes to bolstering the security and reliability of software systems that rely on WASM and native binaries.

In summary, this paper makes the following contributions:
\begin{itemize}
  \item \textbf{First attempt to evaluate WebAssembly decompilers:} 
    As far as we know, we are the first ones to investigate the correctness, readability, and structural similarity of decompilers for WebAssembly. 
    We have created quantifiable and comprehensive metrics, which can be used as useful tools to evaluate the quality of decompiled code. 
    The decompilers are tested on popular benchmarks, synthesized programs, and real-life scenarios to assess their adaptability to various inputs.
  \item \textbf{Inconsistencies of decompiling WASM vs. native binaries:}
    Our investigation delved into the underlying reasons for inconsistencies arising when decompiling WASM vs. native binaries. 
    These observations highlighted several critical issues, including aggressive compiler optimization, WASM language features, and platform-specific concerns. 
  \item \textbf{First analysis framework for decompiled C code:}
    We propose the first comprehensive analysis framework to empirically measure the quality of decompiled C code. 
    We will open-source our benchmark and analysis framework, 
    which can be used for future studies and further advancements in decompilers and WASM analysis.
\end{itemize}

\noindent
The rest of the paper is structured as follows: 
In Section~\ref{sec:motivation}, we present examples of the decompiled code from current WASM decompilers.
In Section~\ref{sec:detail} and Section~\ref{sec:implementation}, we describe the metrics and mechanisms we used to evaluate the decompilers.
The results of our evaluation are presented in Section~\ref{sec:results}. 
In Section~\ref{sec:discussion}, we discuss the limitations of our study and future work.
Finally, we discuss related work in Section~\ref{sec:related} and conclude the paper in Section~\ref{sec:conclusion}.

%% file: Motivation.tex
\section{Motivation}
\label{sec:motivation}

The main goal of this paper is to assess the effectiveness of current decompilers for the WebAssembly (WASM) language and determine their limitations, particularly in their ability to reconstruct source code from WASM precisely. 

We present a motivating example to demonstrate how code readability is enhanced by introducing a decompiler. The example is taken from the paper~\cite{lehmann2020everything} and showcases a stack overflow vulnerability in both C and WASM.

\lstset{frame=tb} 
\lstset{captionpos=t} 
\renewcommand{\lstlistingname}{Listing} 

\begin{lstlisting}[language=C, basicstyle=\footnotesize, caption=C program performing stack overflow (simplified), label={lst:moti_c}]
void vulnerable(char *bar) {
  char buf[8];
  strcpy(buf, bar);  // no bounds checking
}

int main() {
  vulnerable("aaaaaaaaaaaaaaaaaa\04");
}
\end{lstlisting}

\begin{lstlisting}[language=WAT, basicstyle=\footnotesize,caption=Wasm program in WAT (simplified), label={lst:moti_wat}]
(func $f1 (type 14) (param i32)  ;; function vulnerable
    (local i32)
    global.get 0    ;; stack header
    i32.const 16
    i32.sub  
    local.tee 1 
    global.set 0 
    local.get 1  
    i32.const 8
    i32.add         ;; parameter buf
    local.get 0     ;; parameter bar
    call $f2  ;; call function f2 (stpcpy)
    ;; 4 lines omitted
    )  
(func $f2 (type 21) (param i32 i32)
    (local i32)
    block  
        block  
            local.get 0
            ;; 93 lines omitted
            br_if 0 (;@2;)
        end
    end)
\end{lstlisting}

\begin{lstlisting}[language=C, basicstyle=\footnotesize, caption=Decompiled code (simplified) by wasm-decompile, label={lst:moti_wasmdec}]
function f1(a:int) {
  var b:int = g_a - 16;
  g_a = b;
  f2(b + 8, a);
  g_a = b + 16;
}

function f2(a:int, b:int) {
  var c:int;
  if ((a ^ b) & 3) goto B_b;
  // 27 lines omitted
}
\end{lstlisting}

\begin{lstlisting}[language=C, basicstyle=\footnotesize, caption=C program performing summation, label={lst:sum_c}]
int sum(int n) {
  int sum = 0;
  for (int i = 1; i <= n; i++)
    sum += i;
  return sum;
}
\end{lstlisting}

\begin{lstlisting}[language=C, basicstyle=\footnotesize,caption=Decompiled code (simplified) by wasm-decompile, label={lst:sum_dec_wasmdec}]
int f2(int n) {
	if (n <= 0) {return 0; } 
return n - 1 * n - 2 >> 1 + n << 1 - 1}
\end{lstlisting}

The C source code in Listing~\ref{lst:moti_c} contains the \texttt{vulnerable} function, which utilizes the unsafe \texttt{strcpy} function. This is dangerous because \texttt{strcpy} lacks an input size check and can trigger the buffer overflow if \texttt{bar} is larger than \texttt{buf}, which is exactly what Listing~\ref{lst:moti_c} does.

The C source code is compiled into WASM code using Emscripten~\cite{emcc}. WASM code is binary and not human-readable, so it is converted into WAT (WebAssembly Text) format. The WAT code is shown in Listing~\ref{lst:moti_wat}. The \texttt{\$f1} function 
(corresponding to the \texttt{vulnerable} function in the source code, function name striped)
calls the \texttt{\$f2} (\texttt{stpcpy} as aforesaid) function. 
\texttt{\$f2} implements the similar functionality as the \texttt{strcpy} function in the C standard library\footnote{The only difference between \texttt{stpcpy} and \texttt{strcpy} is the return value. \texttt{stpcpy} returns a pointer to the terminating \texttt{\textbackslash0} character of the target string while \texttt{strcpy} returns a pointer to the beginning of the string. Both functions are vulnerable to stack overflow.}. 
The \texttt{\$f2} function has 101 LoC (Lines of Code). 
Since this function is implemented in WASM and does not depend on external functions, one cannot confirm its vulnerability nature without fully understanding it and realizing its similarity to the \texttt{stpcpy} in the C standard library.

We use wasm-decompile~\cite{wabt2_wasm_decompile} to convert the WASM program back into C-like code (Listing~\ref{lst:moti_wasmdec}). The decompiled code has improved readability with 31 LOC (C-like) compared to 101 LOC of WAT. The decreased LoC indicates the improved readability of the program. However, we also need to point out that compared to the original program in Listing~\ref{lst:moti_c}, the decompiled program has increased LoC, and the parameter data type of function \texttt{f1} is different from \texttt{vulnerable} (we would explain this in Section~\ref{sec:results}), which shows the improvement comparing with the WASM binary and limitations regarding the original code.

Besides readability, WASM decompilers may generate incorrect results, which breaks the correctness of the decompiled code.
To demonstrate this, we tested a simple C program that contains a \texttt{sum} function (see Listing~\ref{lst:sum_c}). We compiled the C program into WASM and retained the exported function symbols. We then used wasm-decompile to convert the WASM program into C-like code (Listing~\ref{lst:sum_dec_wasmdec}) and found that the decompiled code was incorrect. Specifically, line 3 in the decompiled code, which calculates the sum from 1 to n, returns $2n-2$ instead of the correct result, $n(n+1)/2$. This could potentially mislead reverse engineers working with the WASM code, leading them to manually check the WASM code instead of the decompiled one to ensure its functionality.

The observed differences between the original and decompiled code raised the following intuitive questions that we aim to address in this paper:

\begin{itemize}
\item\textbf{Q1:} 
To what extent can we trust the decompiled code to present the underlying program's functionality accurately? 
In other words, how do we examine the \textbf{correctness} of the decompiled code? 
Ensuring that the decompiled code faithfully reproduces the original program's functionality is crucial for reliable reverse engineering and code analysis.

\item\textbf{Q2:} 
Some examples of decompiled code (e.g., Listing~\ref{lst:moti_wasmdec}) could still be too long to understand for developers. 
Is there an automated and objective way to measure the \textbf{readability} of the decompiled code? 
Quantifying readability would provide valuable insights into the code's understandability, maintainability, and ease of debugging.

\item\textbf{Q3:} 
Sometimes, even if a piece of code is initially difficult to understand, 
it may still be comprehensible if we can observe a similar structure in the original source code. 
Can we quantify the \textbf{structural similarity} between the decompiled code and the original source code in addition to assessing readability? 
Measuring structural similarity can help determine how closely the decompiled code resembles the original, aiding in code comprehension and validation.

\end{itemize}

Addressing these questions is crucial for advancing the capabilities of decompilers and enhancing the accuracy and usability of decompiled code in various application domains.

%% file: Detail.tex
\section{Methodology}
\label{sec:detail}

\begin{figure*}[h]
\centering
\includegraphics[width=0.8\textwidth]{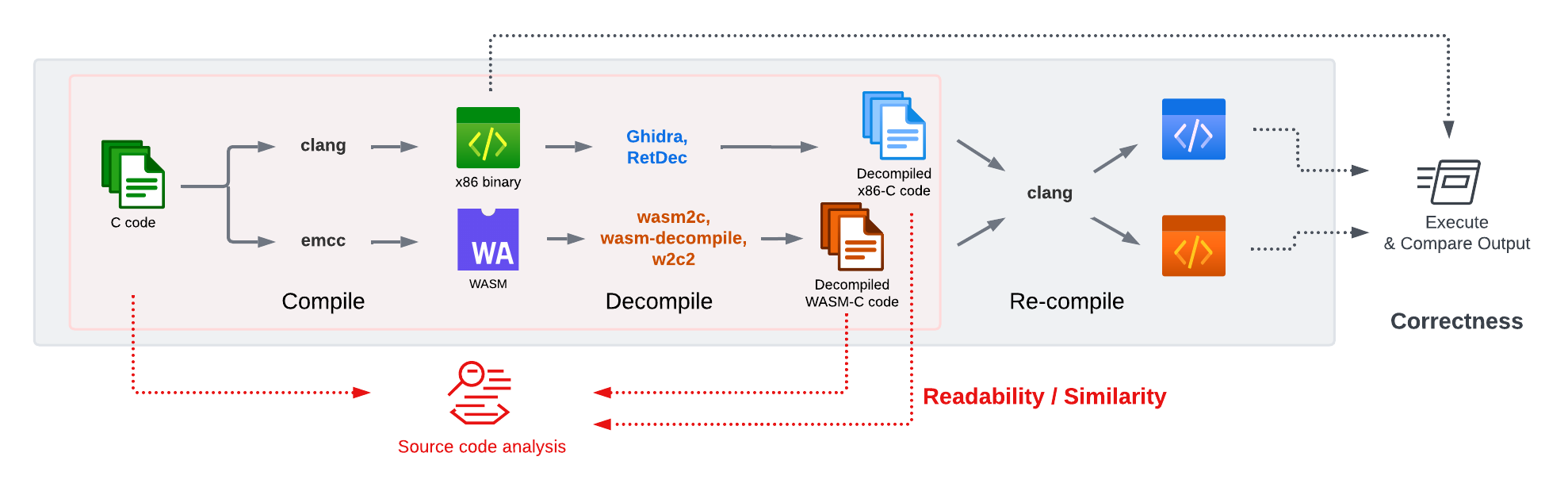}
\vspace{-1em}
\caption{Overview of methodology workflow}
\vspace{-1em}
\label{fig:overview}
\end{figure*}

To answer \textbf{Q1-Q3} from Section~\ref{sec:motivation}, we introduced a series of techniques to comprehensively evaluate existing WASM decompilers in terms of the decompiled output's
correctness, 
readability,
and structural similarity to the original code.

To establish a baseline for comparison, we used the decompiled results obtained from native binary decompilers. We reasoned that these tools have undergone rigorous development and real-world utilization over the years, making them reliable reference points.

To conduct our evaluation, we carefully selected a consistent set of C programs and compiled them separately into both native binaries and WASM files. Subsequently, we applied the chosen decompilers to reverse-engineer the executable files and obtain the decompiled C code from both sides. We then conducted a function-level comparison between the decompiled outputs from native binaries and WASM.

For correctness, we further re-compiled the decompiled code into executable files and executed them to test whether they preserved the same functionality as the original C code.
This step allows us to verify the accuracy of the decompilers in faithfully reproducing the intended behavior of the original programs.

Regarding readability and structural similarity, 
we applied various metrics commonly used in software engineering to evaluate the decompiled code.

An overview workflow of our methodology can be found in Figure~\ref{fig:overview}.

\subsection{Correctness}
\label{subsec:detail_correctness}
The accuracy of decompilation tools plays a crucial role in software development, security, and reverse engineering. 
A perfectly accurate decompiled program should be capable of being re-compiled back into executable files and exhibit the same functionality as the original binary.

However, most state-of-the-art decompilers prioritize enhancing the readability of the decompiled code, 
which often results in C pseudo-code that may not strictly adhere to correct syntax. 
Additionally, these tools typically focus on analyzing the semantics per function and may struggle with reasoning about data structures or global variable access, which are commonly used in C programs.

In short, while readability is a crucial aspect of decompilation, 
accessing the correctness of decompiled code presents two main challenges:
\begin{shaded}
\noindent \textbf{C1: } 
Decompilers often produce C-like code that is not directly re-compilable, 
emphasizing readability over perfect accuracy in reproducing the original binary.

\noindent \textbf{C2: } 
Decompilers may face difficulties effectively handling global variables and memory pointers, which can lead to discrepancies in the decompiled code.
\end{shaded}

On the other hand, WASM decompilers inherently incorporate features that facilitate code re-compilability 
(effectively addressing \textbf{C1}). 
However, challenges persist in generating re-executable programs due to the inherent nature of WASM and its primary use as a library and function within web browsers.
Two properties of WASM introduce divergences in the decompiled code:
\begin{itemize}
\item \textbf{Differences in memory management and representation:} 

WASM employs a stack-based, linear memory model, which contrasts with the memory management and representation used in traditional native binaries. 
This property poses \textbf{C2} as a challenge for WASM decompilers.

\item \textbf{Usage of embedded environment-specific functions:} 

During decompilation, certain operations, such as standard C library functions that access system resources like memory, files, networks, and devices, are transformed into system-call-like functions that exclusively exist within the JavaScript environment of the web browser.
\end{itemize}

The second property of WASM leads to the third challenge for evaluating the correctness of decompiled WASM:
\begin{shaded}
\noindent \textbf{C3: } 
While decompiled WASM programs are represented as C code, 
they commonly cannot be executed directly in a native environment due to the absence of necessary runtime libraries. 
Additionally, 
these decompiled programs faithfully reproduce operations as if they were intended to be used as library modules within web browsers. 
As a result, they lack entry points for direct execution outside the WASM environment, making it infeasible to execute them in a non-WASM context.
\end{shaded}

To address the three challenges (\textbf{C1}, \textbf{C2}, and \textbf{C3}) in evaluating the correctness of decompiled WASM code,
we leveraged the synthesized code generated by DecFuzzer~\cite{liu2020far}
and implemented a sandbox environment for the execution of wasm2c decompiled code.

To the best of our knowledge, DecFuzzer is the only publicly available work that empirically assesses the correctness of native binary decompilers.
It generated code containing only local-variable arithmetic computations into one single ``core function'',
ensuring full decompilability for all existing decompilers (\textbf{C2}).
Subsequently, they syntactically restructured the decompiled code to make it re-compilable (\textbf{C1}). 
The process concluded with the re-compilation of the code, and they compared the output of the re-compiled binary with the original binary's execution results to verify correctness.

As the synthesized programs generated by DecFuzzer do not utilize any library functions or global variables, 
they are suitable for porting to WASM decompilers too. 
We modified them to tackle further \textbf{C2} and \textbf{C3} for WASM.
Specifically, we first transformed the synthesized programs into functionally stand-alone modules. 
This step ensures the compiled WASM is self-contained and does not depend on external runtime libraries and global variables (\textbf{C2}).
Next, we implemented a sandbox environment to load and execute the decompiled WASM code (\textbf{C3}).
This essentially makes it possible to execute the decompiled WASM code in the native binary environment 
so that we can compare it with the original binary's execution results to verify correctness.

We present the detailed implementation in Section~\ref{subsec:implement-correctness}.

\subsection{Readability}
\label{subsec:detail_readability}

\input{tables_metrics_info}

In our study, we introduced several metrics to quantify the readability and complexity of the decompiled programs automatically. 
To assess the complexity of the code, we selected several common algorithms commonly used in software engineering. 
This includes \textit{Lines of code}, 
\textit{Max nesting depth}, 
\textit{Cyclomatic complexity}, and \textit{Halstead complexity measures}.
Generally, the higher the value of these metrics, the more complex and harder to read the code is.
We briefly introduce these metrics below:

\begin{itemize}
\item \textbf{Lines of code.} 
Lines of code measures the total number of lines in the code, indicating the code's size and complexity. 
In our study, we counted only the physical lines in the text of the program's source code, 
excluding comments and blanks \cite{nguyen2007sloc}. 
Generally, a higher number of lines indicates a more complex and potentially harder-to-read code.

\item \textbf{Max nesting depth.}
This metric calculates the maximum depth of nested structures, such as loops (e.g., \codeword{for}, \codeword{while}) and conditionals (e.g., \codeword{if}), within the code. 
A higher value implies deeper nesting, indicating increased complexity and reduced readability~\cite{mcconnell2004code},.

\item \textbf{Cyclomatic complexity.}
Cyclomatic complexity measures a program's control-flow complexity~\cite{mccabe1976complexity}.
It quantifies the number of linearly independent paths through the code, 
representing the number of decision points and possible execution paths. 
Higher cyclomatic complexity values suggest more intricate code structures, making the code more difficult to comprehend.
For implementation, as we are comparing function-to-function instead of the whole program,
we simply count the number of decision points in the function (such as an \codeword{if} statement or \codeword{for} statement).

\item \textbf{Halstead complexity measures.}
The Halstead complexity measures a program's data-flow complexity~\cite{weyuker1988evaluating}. 
The original algorithm includes various metrics such as program vocabulary, program length, volume, difficulty, and effort. 
These measures assess the overall complexity of the code based on the number of distinct operators and operands used and their frequency of occurrence.
In our study, we only calculated program effort for decompiled code. 
\end{itemize}

In addition to the above algorithms, we carefully selected a subset of metrics from previous decompiler works~\cite{yakdan2015no,schulte2018evolving} that are suitable for our function-to-function comparison. 
Specifically,
these works aimed to reduce the reliance on \texttt{goto} statements in the decompiled code, recognizing their potential complexity in comprehension.
Other properties, such as the number of cast and dead assignments, are also evaluated for the presented decompilers.

The full list of metrics used in our research is presented in Table~\ref{tab:metrics_info}. 
We prioritize extensibility and scalability in our metric selection, and to achieve this, we do not include user studies in the scope of our work.
By incorporating these metrics, we aim to comprehensively evaluate the readability and complexity of the decompiled programs, providing valuable insights for future decompiler development and analysis.

\subsection{Structural Similarity}
\label{subsec:detail_similarity}
\begin{figure}[h]%
    \centering
    \subfloat[\small\centering Tree A]{{\includegraphics[width=0.2\textwidth]{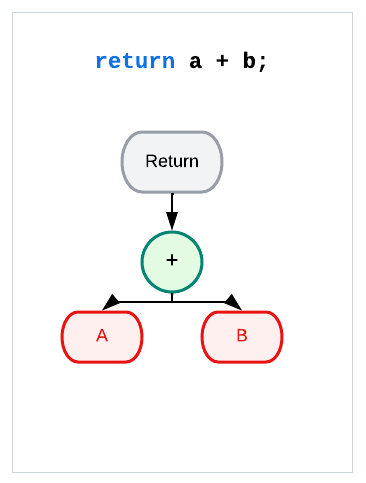} }}%
    \quad
    \subfloat[\small\centering Tree B]{{\includegraphics[width=0.2\textwidth]{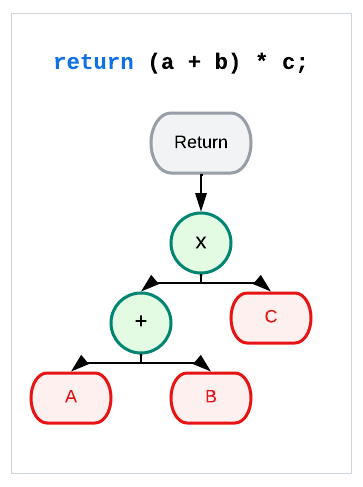} }}%
   \caption{An example of AST comparison}
   \label{fig:ast}
\end{figure}

The Abstract Syntax Tree (AST) is widely used for expressing and analyzing program structures. 
In this paper, we propose utilizing the AST to represent the structure of a program and then compare the AST of the original code with that of the decompiled code to determine their structural similarity.
During the comparison of ASTs, we only consider the ``type'' of each node and not their specific values. 
For instance, in Figure~\ref{fig:ast}, we solely compare the type of each node (represented by colors) and not the actual content of each node (e.g., ``+'', ``*'', ``a'', and ``b'').
The main reason for this design is that we focus only on the structural differences with this metric and leave the numerical evaluation to correctness section.

We employed two comparison algorithms to measure the similarity: 
\textit{Node quantity compare} (NQC) and \textit{Tree edit distance} (TED).

\noindent\textbf{Node quantity compare}. In NQC, we count the number of nodes with the same type between two ASTs. 
The formula to calculate $NQC_{score}$ is shown in Equation~\ref{eq:nqc_formula}. 
The score range falls between $[0, 1]$, 
where a higher score indicates a higher similarity between the two ASTs regarding shared nodes.

\begin{footnotesize}
\begin{equation}%
\sum \frac{common\ \#\ of\ each\ type\ of\ nodes}{max\{total\ nodes\ in\ one\ tree\}}
\label{eq:nqc_formula}
\end{equation}
\end{footnotesize}

Take the two ASTs from Figure~\ref{fig:ast} for example. Among the trees, one $Return$ node, one green node (operation node), and two red nodes (variable nodes) are common. 
The total nodes in Trees A and B are 4 and 6, respectively. 
Therefore, the score of NQC between these two trees is $NQC_{score} = \frac{1+1+2}{max\{4,6\}}\ =\ 0.67$. 
This score indicates a relatively high level of structural similarity between the two trees in terms of common nodes.

\noindent\textbf{Tree edit distance}. 
The \textit{Tree edit distance} (TED) algorithm calculates the number of steps required to transform one tree into another. 
Traditionally, the TED algorithm counts the costs of three operations: \codeword{insert}, \codeword{delete}, and \codeword{replace}.
In our current implementation, we adopt a simplified approach. 
We calculate the differences in depths for each node, which means we only consider the \codeword{insert} and \codeword{delete} operations. 
We do not compare the content of each node during this process.
By focusing solely on the differences in depths and considering only \codeword{insert} and \codeword{delete} operations, 
our implementation provides a straightforward and efficient way to measure the similarity between two trees.

The recursive formula for computing $TED_{score}$ is shown in Equation~\ref{eq:ted_forumla}. Let $F_1 =T_1[i..j]$ be the post-order sub-forest of $T_1$ and let $r_1$ denote its rightmost root. 
Also, let $R_1$ be the rightmost tree of $F_1$ (the one rooted at $r_1$). The same notation holds for $T_2$:

\vspace{0.5em}
\begin{footnotesize}
\noindent\begin{equation}%
\begin{aligned}[t]
 TED&(\emptyset, \emptyset)  =0 \\
 TED&(F_1, \emptyset)  = TED(F_1 - r_1, \emptyset) + cost_{del} \\
 TED&(\emptyset, F_2)  = TED(\emptyset, F_2 - r_2) + cost_{ins} \\
 TED&(F_1, F_2)  = \\& min 
    \begin{cases}
    TED(F_1 - r_1, F_2) + cost_{del} \\
    TED(F_1, F_2 - r_2) + cost_{ins} \\
    TED(F_1 - r_1, F_2 - r_1)  + TED(R_1 - r_1, R_2-r_2)
    \end{cases} \\
 where\ &cost_{del} = cost_{ins} = 1  \\
\end{aligned}
\label{eq:ted_forumla}
\end{equation}
\end{footnotesize}
\vspace{0.5em}

We also normalize and reverse the distance to $TED_{score}$ into values between $[0, 1]$.
$TED(T_1, T_2)$ represents the TED distance between trees $T_1$ and $T_2$, 
while $|T_1|$ and $|T_2|$ denote the total number of nodes in each respective tree.
The larger the $TED_{score}$ we obtain, the more structurally closer the two trees are.

\vspace{0.5em}
\begin{footnotesize}
\noindent\begin{equation}%
   TED_{score} = 1 - \frac{TED(T_1, T_2)}{|T_1|+|T_2|}
\end{equation}
\end{footnotesize}
\vspace{0.5em}

For the example shown in Figure~\ref{fig:ast}, 
with Equation~\ref{eq:ted_forumla},
The TED distance between the two trees is 0.199, resulting in a $TED_{score}$ of 0.801 (1 - 0.199). 
This score indicates a relatively high structural similarity between the two trees.

\subsection{Compilers \& Decompilers}
\label{sec:com_decom_intro}

In our C-to-WASM compilation process, we opted to use \textit{Emscripten} (\textit{emcc})~\cite{emcc}, 
a comprehensive compiler toolchain for WASM built on \textit{LLVM}~\cite{llvm}. 
As emcc utilizes LLVM as its foundation, 
we chose to employ \textit{clang} to generate the native binary to ensure consistency in both compilation and re-compilation. 
Additionally, we aimed to set the compilation options to be identical to those used within the emcc implementation.

For the WASM decompilation process, we utilized three widely used WASM decompilers: \textit{wasm2c}~\cite{wabt1_wasm2c}, \textit{w2c2}~\cite{w2c2}, and \textit{wasm-decompile}~\cite{wasmdec}. 

Wasm2c is part of the WABT WebAssembly tool set~\cite{wabt}. 
It converts a WASM binary file to C source code along with the auxiliary header and supports various experimental WASM features, such as exceptions and threading, by incorporating specific command-line options. Wasm2c can generate bug-free C programs that can be compiled and executed, and the behavior of the recompiled program is expected to adhere to the original program; that is, the correctness of the program shall hold.

Wasm-decompile~\cite{wasmdec} is also a component of WABT toolset. The reason to include another tool from the WABT toolset is 
that wasm-decompile and wasm2c have very different design goals, and the difference is significant enough to include both decompilers in our study (see Sec.~\ref{sec:results}). Wasm-decompile ``is aimed at users that want to be able to `read' large volumes of Wasm code~\cite{wasmdec}." Decompiled code generated by {wasm-decompile is not designed to "be a programming language," that is, the functionality of recompilation is currently not provided, and the execution correctness of the decompiled code is not guaranteed.
In conclusion, wasm-decompile focuses on the decompiled code's readability and not correctness. 

W2c2~\cite{w2c2} is a standalone tool that translates WASM modules to portable C. 
While it supports basic WebAssembly features, it also includes three experimental WASM features: bulk memory operations, sign-extension operators, and non-trapping float-to-int conversions.

In the case of native binary decompilation, 
we employed two open-source decompilers that were previously evaluated in the DecFuzzer paper: 
\textit{Ghidra}~\cite{ghidra} and \textit{RetDec}~\cite{retdec}. 
These tools have been widely used and evaluated in various decompilation tasks, 
making them suitable candidates for our evaluation.

\subsection{Benchmarks}
\label{sec:benchmarks}
We use the 1,000 CSmith synthesized C programs from DecFuzzer for correctness.
In the original work of DecFuzzer, the authors further mutated them to test decompilers rigorously.
We considered 1,000 programs reasonable for our attempt to examine WASM decompilers, and we did not mutate them further.

For evaluating structural similarity and readability, we employed two widely used C benchmark suites: 
PolyBenchC~\cite{polybench/c} and CHStone~\cite{chstone}. 
These benchmark suites contain real-world program implementations that are likely to be used in the WASM environment, 
such as scientific visualization, encryption, simulation, image recognition, etc.

We opted not to include the programs from DecFuzzer for structural similarity and readability evaluation for two main reasons. 
Firstly, we found that since DecFuzzer's synthesized programs may contain dead code, some portions of the code are automatically removed during compiling, even when setting the compiler optimization level to 0. This results in the decompilers not handling the exact same program as the original one, making it less reasonable to include these programs.

Secondly, the original synthesized code from DecFuzzer is not designed for human readers and contains no specific meaning or intended functionality. 
Consequently, including these programs in the evaluation would not be meaningful for assessing decompiled code's structural similarity and readability.

In contrast, PolyBenchC and CHStone are not used in the correctness evaluation due to their inclusion of library calls, which are challenging for both native and WASM decompilers to handle correctly during runtime. 
Therefore, we reserved these benchmark suites to assess the structural similarity and readability aspects of the decompiled programs.

%% file: tables_metrics_info.tex
\begin{table*}[ht]
  \centering
  \caption{List of readability and structural similarity metrics}
\resizebox{0.85\textwidth}{!}{%
\small
  \begin{tabular}{|p{0.25\linewidth}|p{0.525\linewidth}|p{0.075\linewidth}|}
    \hline
      \textbf{Metric} & \textbf{Definition} & \textbf{Range} \\  
    \hline
     Line of Code & 
      Source Lines of Code & 
      $[0, \infty)$ \\
      Max Nesting Depth &
      Depth of the most nested statement &
      $[0, \infty)$\\
      Cyclomatic Complexity &
      \# of linearly independent paths (Control-Flow Complexity) &
      $[1, \infty)$ \\
      Halstead Complexity Measure &
      Score proportional to operators and operands (Data-Flow Complexity) &
      $[0, \infty)$ \\
      \# of goto, \# of variables, \newline
      \# of cast, \# of dead assignments &
      \multirow{2}{*}{-} &
      \multirow{2}{*}{$[0, \infty)$} \\
    \hline
     \hline
     AST Node Quantity Compare & 
      Common types of nodes between two ASTs & 
      $[0, 1]$ \\
     AST Tree Edit Distance & 
      Tree edit distance between two ASTs & 
      $[0, 1]$ \\
     \hline
  \end{tabular}
}

\label{tab:metrics_info}
\end{table*}

%% file: Implementation.tex
\section{Implementation}
\label{sec:implementation}

The total scripts for compiling benchmarks and generating metrics contain around 5,000 lines of Python and shell scripts.

\subsection{Correctness}
\label{subsec:implement-correctness}

DecFuzzer utilized CSmith \cite{csmith}, a widely recognized C-code synthesizer for compiler testing, to generate C code for evaluating decompilers. 
In its original form, the generated code operated on global variables and computed a checksum to verify execution results.
DecFuzzer consolidated the synthesized code into a single function, \codeword{func}, and operated on local copies of the original global variables.
As \codeword{func} contains no library calls or global variable access, it can be fully compiled and decompiled by state-of-the-art decompilers.
To generate re-compilable code, 
DecFuzzer deployed a simple rewriter to fix the syntax errors in decompiled code.
Finally, the rewritten code was re-compiled and executed to test against the original binary.

To adapt DecFuzzer for WASM, we wrapped \codeword{func} into an importable module that can be called and receive output from the main function. 
This module-based approach was chosen because the WASM code handles memory differently, 
making it challenging to create a complete program that can be executed natively. 
Luckily, as a standalone WASM module, the code can be imported and called within a testing framework simulating a browser environment.
This modified code was then compiled to WASM, decompiled back to C, re-compiled to object files, 
and eventually imported and called by our sandbox testing framework.

Besides the above efforts, we upgraded DecFuzzer's syntax rewriters to match the latest compiler and decompiler versions. 
We also switched from using GCC to clang, 
which aligned better with the WASM toolchain and resulted in fewer errors during preliminary testing. 
To make the DecFuzzer code compatible with WASM, 
we created a Python converter with approximately 300 lines of code. 
Additionally, we patched the legacy RLBox\cite{wasmboxc} code to match the latest WASM toolchain and integrated it into our evaluation framework for sandbox testing.
Besides,
\textit{w2c2} is skipped for our correctness evaluation due to the current lack of support for sandboxing the \textit{w2c2} decompiled code.

\subsection{Readability \& Structural Similarity}
\label{subsec:imp_read_ss}
For readability and structural similarity, we built the analysis on top of \textit{clang Python API}~\cite{clang_python} and \textit{cppcheck}~\cite{cppcheck}.
As \textit{Clang API} relies on the Clang preprocessor, the under-analyzing target must be C syntactically correct.
We leveraged the syntax rewriter of DecFuzzer to try to make the native binary's decompiled code re-compilable. 
Due to the limitation of the preprocessor, it cannot parse the decompiled code generated by wasm-decompile under optimization level 1 or 2, as the code would break the type system during the analysis progress. 

%% file: Evaluation.tex
\section{Evaluation Results}
\label{sec:results}

As introduced in the previous chapters, 
we evaluated the state-of-the-art WASM decompilers, \textit{w2c2}, \textit{wasm2c}, and \textit{wasm-decompile}, in three
aspects: correctness, readability of the decompiled code, 
and structural similarity of the decompiled code with the source code. 
Based on decompiler characteristics introduced in Section~\ref{sec:com_decom_intro}, w2c2 is excluded from the correctness evaluation, and the structural similarity evaluation of wasm-decompile is limited.

To establish a baseline for comparison, 
we also included decompilers of native binaries, namely \textit{Ghidra} and \textit{RetDec}.
All experiments were performed on an Intel-i5 machine with four cores, 8GB RAM, and running ArchLinux. 
To ensure the validity of the results, we used the latest versions of all software.

For the selected benchmarks, we observed that the compiling and decompiling processes could be completed within five seconds without significant memory consumption. 
Given the swift execution of these processes, we did not measure and compare the decompilers' runtime performance. 
Instead, our focus was solely on evaluating the quality of the generated code.

In each section of the evaluation, we address the following questions through the use of proposed metrics:
\begin{itemize}
   \item Can proposed metrics help evaluate and characterize decompilers?
   \item How does each decompiler perform in terms of the specific metrics?
\end{itemize}

\subsection{Correctness}
\label{subsec:result_correctness}

Table~\ref{tab:correctness_results} shows the result of our correctness evaluation.
Factors are based on DecFuzzer, including execution results and the number of decompile or re-compile failures.

From the table,
we can clearly see that \textit{wasm2c} outperformed native binary decompilers by
achieving a 100\% correctness rate.
We attribute this success to the fact that \textit{wasm2c} adopts a conservative approach when translating WASM to decompiled code: 
Performing minimal optimizations. 
Consequently, when the execution environment was appropriately set up, the decompiled code faithfully reproduced the original functionality.
This high correctness rate can also be attributed to the more uncomplicated instruction set in WASM compared to native binary code.
Therefore, it's relatively simple to translate WASM bytecode to C code directly,
whereas native binary decompilers often face challenges in handling low-level instructions. 
Propagating these instructions as functions in the decompiled code without providing corresponding runtime implementations often leads to re-compile failures.

\input{tables_correctness_results.tex}

Compared to \textit{wasm2c}, the other three decompilers all suffer from compilation failure and execution discrepancies. 
First, as presented to be a decompiler to improve readability for WASM, 
\textit{wasm-decompile} is only able to achieve 649 out of 1000 program correct.
We manually reviewed the semantically incorrect code, and summarized the reason for failures into two main reason:

\noindent \textbf{Unfaithful representation of the original WASM workflow}
In order to improve redability,
\textit{wasm-decompile} translates WASM's stack-based instructions into 
SSA formats, which is theoretically more readable for human
than long lines of C code in \textit{wasm2c}.
However, this transformation also introduces errors in the 
value passing and data flow, which leads to the incorrect execution of the decompiled code.

\noindent \textbf{False identification of structures and pointers}
In the synthesized benchmark for testing correctness, 
we introduce no pointer or special data structure to reduce the complexity of evaluation.
However, \textit{wasm-decompile} falsely grouped some of the variables as data structures 
and accessed through pointer offsets on stack.
This not only results in violation of C syntax that resulted in incompilabiity,
but potentially introduce rooms for more error to preserve the original program's
semantic.
Moreover, it conversely increases the difficulty for understanding the decompile code.

We showed a code snippet of the decompile output for \textit{wasm-decompile}
in Listing~\ref{lst:d_wasm-decompile_h}.
The first part of the code is how \textit{wasm-decompile} translates
function epilogue. 
As all function parameters are stored on stack and then loaded during function calls,
\textit{wasm-decompile} may falsely group some parameters as one data structure, 
while the original program actually only passes integer values.
The latter part of the listing shows how variable type casting is represented with 
\textit{wasm-decompile}. 
The variable \texttt{n} is a 32-bit integer casting to 8-bit and stored into
\texttt{q}. 
This kind of operation sequence is conducted through out the decompile code,
making the code not only long and hard to read,
but also introducing correctness errors throughout the long dependency chain.

{\renewcommand\thelstnumber{%
g\arabic{lstnumber}\fi}}
\begin{lstlisting}[language=C, basicstyle=\footnotesize, caption=Code snippet of decompile code generated by wasm-decompile, label={lst:d_wasm-decompile_h}]
export function func_1():int {
  var a:int = stack_pointer;
  var b:int = 16;
  var c:{ a:int, b:int /*...*/ } = a - b;
  stack_pointer = c;
  
  var f:int = -1051244671;
  c.b = f;
  var g:int = 1902055121;
  c.a = g;
  // ....
  var o:int = 24;
  var p:int = n << o;
  var q:int = p >> o;
  // ....
}
\end{lstlisting}

For native-binary decompilers,
both \textit{Ghidra} and \textit{RetDec} achieved slightly better results than \textit{wasm-decompile},
but still have a significant number of failures.
The main reason for the failures is the decompilers' 
capabilities of translating low-level instructions to high-level C code.
For such instructions, the decompilers often directly translate them to function calls, mimicking the original instructions.
The runtime environment does not support such cases
and will either lead to re-compile failures 
or be falsely rewritten by the DecFuzzer, leading to incorrect execution results.
For example, \textit{Ghidra} 
places \codeword{CONCAT} as a pseudo function to translate assembly code instructions such as \texttt{mov AH, 2}.
While serving a purpose in the original code,
the existing DecFuzzer implementation
falsely removed it from the decompiled code to make the decompiled program re-compilable.

\begin{shaded}
   \noindent\textbf{Summary:}
   With the presented approach for evaluating correctness,
   we highlight the remarkable 100\% accuracy of \textit{wasm2c}'s decompiled code. 
   The limitation of \textit{wasm-decompile} in correctness not only
   emphasizes the importance of evaluating decompilers from multiple perspectives 
   but also underscores the challenges in recovering WASM's stack-based instructions into high-level C code.
   While native binary decompilers present some failures, 
   it is essential to note that some may be attributed to the evaluation approach rather than inherent issues with the decompilers themselves. 
   More refined evaluation methodologies and techniques can address and improve these specific cases separately.
\end{shaded}

\input{tables_metrics_results.tex}

\input{tables_metrics_plots.tex}

\subsection{Readability}
\label{subsec:result_readability}

We conducted a comprehensive evaluation by compiling each C code in PolyBenchC and CHStone at optimization levels 0, 1, and 2 to both WASM and native binaries and subsequently decompiled outputs to compare their differences.

The complete evaluation results are in Table~\ref{tab:readability_results} and Figure~\ref{fig:results_plots}. Regarding the readability metrics, higher values generally indicate more challenging to understand.

First, by comparing the results for \textit{Lines of code}, \textit{Maximum nesting depth}, \textit{Cyclomatic complexity}, and \textit{Halstead complexity measure}, it is evident that WASM decompilers generated significantly different code compared to others. 
The decompiled WASM code contains more lines of code and variables. 
Structurally, it exhibits a ``flattened'' code with a relatively low maximum nesting depth and Cyclomatic complexity. 
However, as the Halstead complexity measure indicates, the data flows are notably complicated. 
The low nesting depth also comes from the fact that WASM's decompiled code relies heavily on goto statements.
This discrepancy is mainly due to the fundamental differences between WASM's stack-based instruction set and C's register-based language. 
The state-of-the-art native binary decompilers attempt to translate the code to resemble the semantics of C code closely, 
while the WASM decompilers lack such design as observed in the motivation decompiled code (Section~\ref{sec:motivation}), all three WASM decompilers primarily convert WASM's stack-based instructions into C syntax directly.

Although \textit{wasm-decompile} is designed to generate readable decompiled code,
it is interesting that the readability metrics do not necessarily support this statement.
With optimization level 0, decompiled code generated by \textit{wasm-decompile} has a significantly larger number
in most metrics evaluated because \textit{wasm-decompile} attempts to recover the WASM semantics into readable format faithfully. 
However, as WASM has only a limited number of types, 
a large portion of the decompiled code is related to type conversion between variables in support of the original C programs' semantics.
Faithfully translating these operations could help to understand small-scale analysis of variable dependency better,
but may not necessarily improve the whole program's readability.
This is highly reflected in the differences in the total LoC and type castings between \textit{wasm-decompile} and \textit{wasm2c}.
\textit{wasm2c} does not specify the casting of variables in code while preserving the correct semantics as shown in
Section~\ref{subsec:result_correctness},
while \textit{wasm-decompile} explicitly shows such program flow and resulted in higher numbers in these two metrics.

In comparing native binary decompilers, \textit{Ghidra} generally generated more compact code and was closer to the original. 
This is attributed to \textit{Ghidra}'s continuous development and successful heuristics in decompiling. 
In contrast, \textit{RetDec} produced more lines of code and variables. This is due to its failures in recovering many native binary instructions into high-level C code and simply migrating instructions to C functions.
E.g., \textit{RetDec} directly used \codeword{v8 = \_\_asm\_movsd(v3)} in the decompiled code as the translation of the instruction \codeword{movsd}.

For other metrics, some may not directly reflect the code's readability but offer insights into the characteristics of the generated decompiled code. 
For instance, \textit{Ghidra} had excessive type casting.
This is due to its strategy for pointers and data-structure recovery, but it may not essentially affect how we interpret the code.

Interestingly, higher compilation optimization levels have different effects on different decompilers. 
WASM decompilers tend to produce more compact code at higher optimization levels, 
mainly because highly optimized WASM files are smaller in size, resulting in shorter decompiled code when directly translated. 
However, this does not necessarily mean that the decompiled code more closely resembles the original program nor
improve readability.
For example, 
in Listing~\ref{lst:wasm-decompile_opt2}, we show a code snippet of decompile code genereated by
\textit{wasm-decompile} with optimization level 2.
It is observable that although the total lines of code could be less as operations are 
shrinking into single lines, but the code is syntactically incorrect and semantically incomprehensible
for the complicated value assignments between operations.

\begin{lstlisting}[language=C, basicstyle=\footnotesize, caption=Decompiled code by wasm-decompile with compiler optimization, label={lst:wasm-decompile_opt2}]
export function KeySchedule(a:int_ptr, /*...*/):int {
    // ... 
    c = b - (g = b / i) * i;
    if (eqz(c)) {
        f = Sbox;
        c = f + ((e = (c = ((a = word + (a << 2)) + 1440)[0]:int) / 16) << 6) + (c - (e << 4) << 2);
        e = f + ((d = (e = a[240]) / 16) << 6) + (e - (d << 4) << 2);
        h = (f + ((h = (d = a[120]) / 16) << 6))[d - (h << 4)]:int ^
            (Rcon0 + (g << 2) - 4)[0]:int;
        g = a[0];
      goto B_q;
    }
    // ....
}
\end{lstlisting}

On the contrary, higher optimization levels do not necessarily lead to more compact code for native binary decompilers. 
Native compilers use aggressive algorithms to optimize binary size and performance, 
while these optimized operations may not be easily translated into C semantics. 
Decompilers typically create highly readable code through heuristic approaches. 
When a binary is highly optimized, it may not fit within the existing heuristics, 
resulting in larger decompiled code with more direct assembly-to-C translations.

\begin{shaded}
   \noindent\textbf{Summary: }
   The evaluation results unveil substantial differences in the decompiled code generated by WASM decompilers. 
   Notably, the decompiled code exhibits excessive lines of code, frequent usage of goto statements, and intricate data flows, all of which negatively impact code readability.

   Moreover, these metrics provide valuable insights into other interesting aspects, such as the influence of compiler optimization levels on each decompiler's performance. 
   The interplay between these factors can be considered to be a robust indicator for assessing the quality of decompiled code generated by the decompilers.
\end{shaded}

\subsection{Structural Similarity}

The evaluation of structural similarity yields similar observations to the readability metrics. 
The results of structural similarity can be found in Table~\ref{tab:similarity_results}. 
As we have mentioned before, due to parser limitation, \textit{wasm-decompile}-generated code can only be analyzed under optimization level 0.
The evaluation compared original and decompiled functions by normalizing the computed \textit{Node Quantity Compare} (NQC) and \textit{Tree Edit Distance} (TED) scores to a range of $[0,1]$. 
Higher scores indicate greater structural similarity between the functions.

In general, the structural similarity results align with the findings from the readability evaluation. 
WASM decompiled code exhibits significant differences compared to native binary decompiled code, and the trend of structural similarity varies with different optimization levels. 
Higher optimization levels lead to increased structural similarity for WASM decompilers, while for native binary decompilers, the opposite trend is observed.

Notably, the structural similarity evaluation revealed insights that are not evident in the readability evaluation. 
Specifically, the improvements introduced by \textit{w2c2} are more apparent. 
We observed several exceptional high scores for \textit{w2c2} decompiled code, corresponding to \textit{w2c2}'s outliers in Figure~\ref{fig:results_plots}e and Figure~\ref{fig:results_plots}f. 
Our manual inspection revealed that these scores are primarily associated with pure arithmetic functions like \codeword{mul64To128}, which do not contain pointers or data structures. 
These functions exhibit minimal structural differences between the compiled WASM and native binary. 
As a result, \textit{w2c2} successfully decompiled them with high NQC and TED scores that are closer to what native binary decompilers obtained. In contrast, \textit{wasm2c} still produced long and complicated code for these functions.

\begin{shaded}
   \noindent\textbf{Summary: }
   Overall, the structural similarity evaluation provides additional insights into the performance of decompilers.
   Specifically, it highlights the effectiveness of \textit{w2c2} in handling certain types of functions that exhibit minimal structural differences between compiled WASM and native binary versions.
\end{shaded}

%% file: tables_correctness_results.tex
\begin{table}[h]
\centering
\caption{Correctness results.}
\resizebox{\columnwidth}{!}{%
\small
\begin{tabular}{|l|ccc|c|} 
\hline
\textbf{Bench} & \textbf{Success} & \textbf{Re-compile Failure} & \textbf{Exec. Discrepancy} & \textbf{Total} \\
\hline
   wasm2c & 1,000 & 0 & 0 &  \\
   wasm-decompile & 649 & 6 & 345 & \\
   Ghidra & 818 & 13 & 169 & 1,000\\
   RetDec & 775 & 0 & 225 & \\
\hline
\end{tabular}
}

\label{tab:correctness_results}
\end{table}

%% file: tables_metrics_results.tex
\begin{table*}[ht]
\centering
\caption{Results of readability evaluation}
\resizebox{0.95\textwidth}{!}{%
\small
\begin{tabular}{|l|c|cc|cc|cc|cc|cc|cc|}
\hline
\multirow{2}{*}{\textbf{Metrics}} & \multirow{2}{*}{\textbf{Opt Level}} & \multicolumn{2}{c|}{\multirow{2}{*}{\textbf{Original Code}}} & \multicolumn{10}{c|}{\textbf{Decompilers}}  \\
 \cline{5-14}
&  & \multicolumn{2}{c|}{} & \multicolumn{2}{c|}{\textbf{w2c2}} & \multicolumn{2}{c|}{\textbf{wasm2c}}  & \multicolumn{2}{c|}{\textbf{wasm-decompile}} & \multicolumn{2}{c|}{\textbf{Ghidra}} & \multicolumn{2}{c|}{\textbf{RetDec}}\\
\hline
\hline
 &  & \multicolumn{12}{c|}{Total} \\
\hline
 & O0 & \multicolumn{2}{c|}{} & \multicolumn{2}{c|}{74,473} & \multicolumn{2}{c|}{72,362} & \multicolumn{2}{c|}{23,744} & \multicolumn{2}{c|}{10,376} & \multicolumn{2}{c|}{4,818}  \\
Lines of code & O1 & \multicolumn{2}{c|}{2,619} & \multicolumn{2}{c|}{24,828} & \multicolumn{2}{c|}{28,332} & \multicolumn{2}{c|}{7,975} & \multicolumn{2}{c|}{11,290} & \multicolumn{2}{c|}{7,487} \\
 & O2 & \multicolumn{2}{c|}{} & \multicolumn{2}{c|}{27,440} & \multicolumn{2}{c|}{30,461} & \multicolumn{2}{c|}{7,771} & \multicolumn{2}{c|}{15,575} & \multicolumn{2}{c|}{9,249}  \\
\hline
 & O0 & \multicolumn{2}{c|}{} & \multicolumn{2}{c|}{767} & \multicolumn{2}{c|}{576} & \multicolumn{2}{c|}{768} & \multicolumn{2}{c|}{98} & \multicolumn{2}{c|}{21}  \\
\# of goto statements & O1 & \multicolumn{2}{c|}{0} & \multicolumn{2}{c|}{738} & \multicolumn{2}{c|}{296} & \multicolumn{2}{c|}{774} & \multicolumn{2}{c|}{175} & \multicolumn{2}{c|}{63}  \\
 & O2 & \multicolumn{2}{c|}{} & \multicolumn{2}{c|}{584} & \multicolumn{2}{c|}{257} & \multicolumn{2}{c|}{455} & \multicolumn{2}{c|}{405} & \multicolumn{2}{c|}{131}  \\
\hline
 & O0 & \multicolumn{2}{c|}{} & \multicolumn{2}{c|}{0} & \multicolumn{2}{c|}{0} & \multicolumn{2}{c|}{627} & \multicolumn{2}{c|}{138} & \multicolumn{2}{c|}{2,872}  \\
\# of type casting & O1 & \multicolumn{2}{c|}{39} & \multicolumn{2}{c|}{0} & \multicolumn{2}{c|}{0} & \multicolumn{2}{c|}{382} & \multicolumn{2}{c|}{99} & \multicolumn{2}{c|}{2,520}  \\
 & O2 & \multicolumn{2}{c|}{} & \multicolumn{2}{c|}{0} & \multicolumn{2}{c|}{0} & \multicolumn{2}{c|}{642} & \multicolumn{2}{c|}{92} & \multicolumn{2}{c|}{2,958}  \\
\hline
 & O0 & \multicolumn{2}{c|}{} & \multicolumn{2}{c|}{16,109} & \multicolumn{2}{c|}{19,368} & \multicolumn{2}{c|}{18,804} & \multicolumn{2}{c|}{2,883} & \multicolumn{2}{c|}{519}  \\
\# of variables & O1 & \multicolumn{2}{c|}{194} & \multicolumn{2}{c|}{1,496} & \multicolumn{2}{c|}{1,591} & \multicolumn{2}{c|}{619} & \multicolumn{2}{c|}{3,261} & \multicolumn{2}{c|}{760}  \\
 & O2 & \multicolumn{2}{c|}{} & \multicolumn{2}{c|}{1,622} & \multicolumn{2}{c|}{1,711} & \multicolumn{2}{c|}{741} & \multicolumn{2}{c|}{4,952} & \multicolumn{2}{c|}{1,110}  \\
\hline
 & O0 & \multicolumn{2}{c|}{} & \multicolumn{2}{c|}{64} & \multicolumn{2}{c|}{207} & \multicolumn{2}{c|}{7} & \multicolumn{2}{c|}{4} & \multicolumn{2}{c|}{31}  \\
\# Lines of dead code & O1 & \multicolumn{2}{c|}{4} & \multicolumn{2}{c|}{143} & \multicolumn{2}{c|}{25} & \multicolumn{2}{c|}{14} & \multicolumn{2}{c|}{4} & \multicolumn{2}{c|}{7}  \\
 & O2 & \multicolumn{2}{c|}{} & \multicolumn{2}{c|}{513} & \multicolumn{2}{c|}{140} & \multicolumn{2}{c|}{35} & \multicolumn{2}{c|}{8} & \multicolumn{2}{c|}{8}  \\
\hline
\hline
 &  & Average & Stdev & Average & Stdev & Average & Stdev & Average & Stdev & Average & Stdev & Average & Stdev  \\
\hline
 & O0 & & & 1.221 & 1.534 & 0.655 & 0.477 
 & 1.455 & 1.409 & 1.634 & 1.798 & 1.338 & 1.464  \\
Maximum nesting depth & O1 & 0.952 & 1.163 & 1.228 & 1.558 & 0.600 & 0.492 
& 1.041 & 1.241 & 1.855 & 2.010 & 2.000 & 1.911 \\
 & O2 &  &  & 1.766 & 2.447 & 1.462 & 1.514 
 & 1.710 & 2.065 & 2.097 & 2.428 & 1.745 & 1.813 \\
\hline
& O0 & & & 3.634 & 3.853 & 4.924 & 4.667 & 6.283 & 6.139 & 6.772 & 8.465 & 4.683 & 4.412 \\
Cyclomatic complexity & O1 & 14.329 & 18.067 & 5.152 & 7.295 & 6.766 & 8.030  & 6.710 & 8.475 & 7.517 & 9.430 & 7.883 & 8.920 \\
& O2 & & & 5.883 & 8.192 & 7.359 & 8.530 & 6.228 & 8.180 & 9.255 & 12.438 & 9.048 & 10.252 \\
\hline
 & O0 & & & 43.177 & 31.158 & 52.026 & 24.492 & 26.084 & 9.363 & 30.956 & 30.761 & 36.106 & 44.118 \\
Halstead complexity measure & O1 & 3.996 & 4.144 & 46.341 & 54.848 & 52.563 & 51.660 & 30.778 & 26.185 & 31.434 & 26.192 & 35.865 & 36.894 \\
 & O2 &  &  & 55.253 & 62.667 & 66.963 & 63.842 & 36.559 & 32.038 & 30.818 & 25.237 & 42.948 & 42.853 \\
\hline
\end{tabular}
}

\label{tab:readability_results}
\end{table*}

\begin{table*}[ht]
\centering
\caption{Results of structural similarity evaluation}
\resizebox{0.85\textwidth}{!}{%
\small
\begin{tabular}{|l|c|cc|cc|cc|cc|cc|}
\hline
\multirow{2}{*}{\textbf{Metrics}} & \multirow{2}{*}{\textbf{Opt Level}} & \multicolumn{10}{c|}{\textbf{Decompilers}}  \\
 \cline{3-12}
& & \multicolumn{2}{c|}{\textbf{w2c2}} & \multicolumn{2}{c|}{\textbf{wasm2c}} & \multicolumn{2}{c|}{\textbf{wasm-decompile}} & \multicolumn{2}{c|}{\textbf{Ghidra}} & \multicolumn{2}{c|}{\textbf{RetDec}} \\
\hline
\hline
 & & Average & Stdev & Average & Stdev & Average & Stdev & Average & Stdev & Average & Stdev  \\
\hline
                      & O0 & 0.170   & 0.206 & 0.047   & 0.022 & 0.134   & 0.066 & 0.573   & 0.260 & 0.506   & 0.310 \\ 
Node quantity compare & O1 & 0.230   & 0.195 & 0.197   & 0.140 & N/A     & N/A   & 0.470   & 0.284 & 0.378   & 0.291 \\
                      & O2 & 0.223   & 0.201 & 0.193   & 0.147 & N/A     & N/A   & 0.436   & 0.304 & 0.357   & 0.300 \\
\hline
                      & O0 & 0.166   & 0.169 & 0.062   & 0.046 & 0.149   & 0.103 & 0.495   & 0.236 & 0.439   & 0.286 \\ 
Tree edit distance    & O1 & 0.219   & 0.158 & 0.184   & 0.151 & N/A     & N/A   & 0.403   & 0.236 & 0.343   & 0.257 \\
                      & O2 & 0.210   & 0.166 & 0.191   & 0.152 & N/A     & N/A   & 0.377   & 0.256 & 0.325   & 0.267 \\
\hline
\end{tabular}
}
\label{tab:similarity_results}
\end{table*}

%% file: tables_metrics_plots.tex
\begin{figure*}[h]%
    \centering
    \subfloat[\small\rmfamily\centering Line of Code]{{\includegraphics[width=0.30\textwidth]{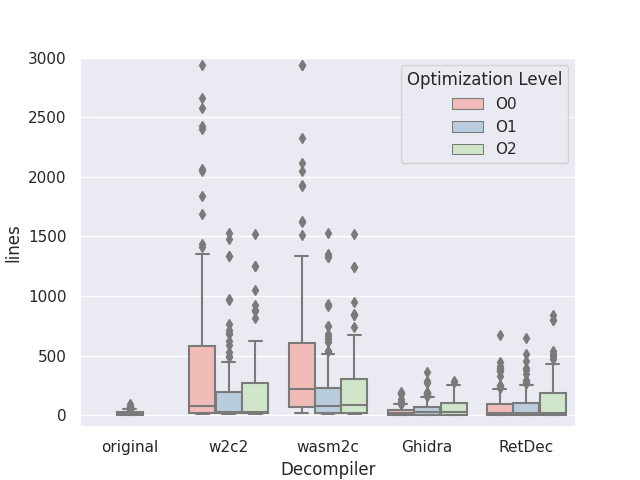} }}%
    \quad
    \subfloat[\small\rmfamily\centering Maximum Nesting Depth]{{\includegraphics[width=0.30\textwidth]{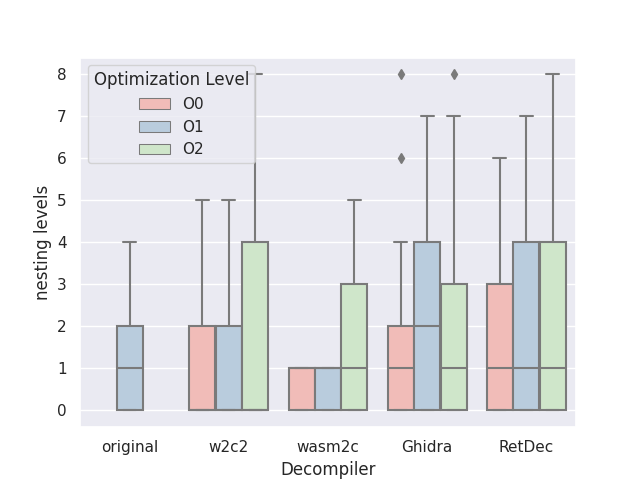} }}%
    \quad
    \subfloat[\small\rmfamily\centering Cyclomatic Complexity]{{\includegraphics[width=0.30\textwidth]{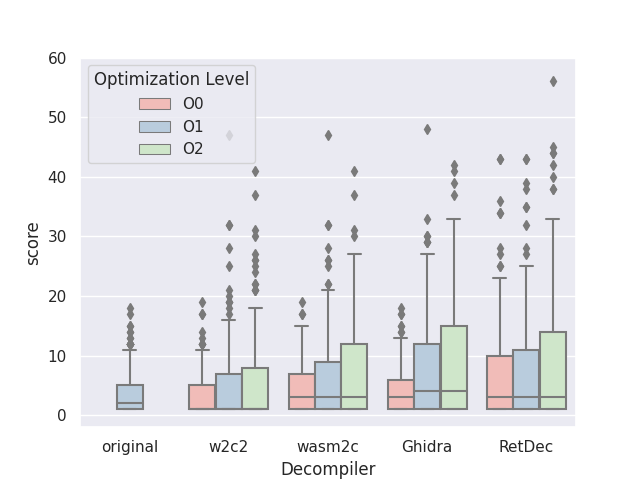} }}%
    \quad
    \subfloat[\small\rmfamily\centering Halstead Complexity Measure]{{\includegraphics[width=0.30\textwidth]{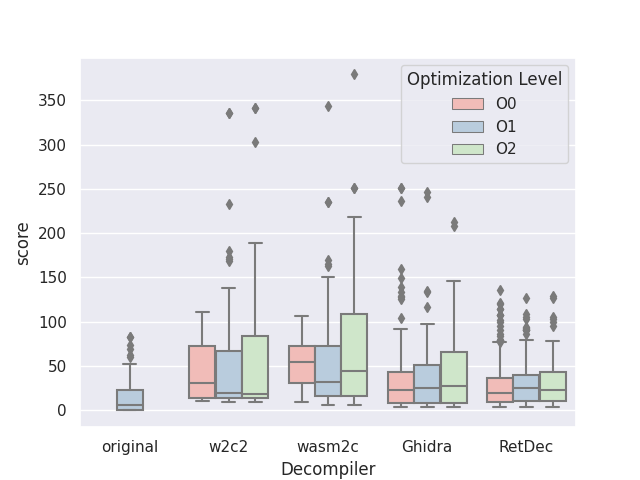} }}%
    \quad
    \subfloat[\small\rmfamily\centering Node Quantity Compare ]
    {{\includegraphics[width=0.30\textwidth]{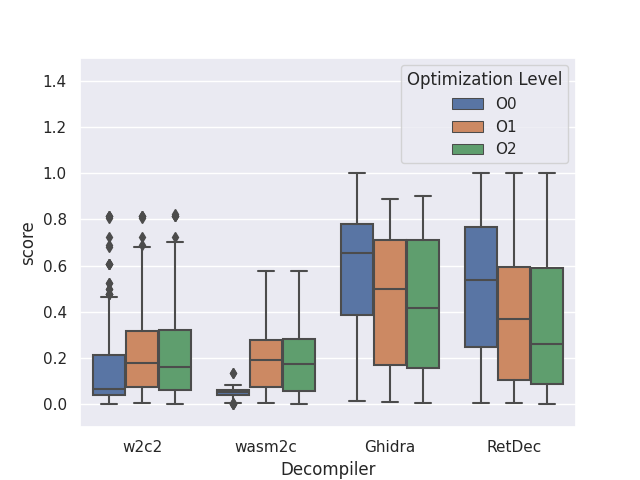} }}%
    \label{fig:plot_nqc}
    \quad
    \subfloat[\small\rmfamily\centering Tree Edit Distance]
    {{\includegraphics[width=0.30\textwidth]{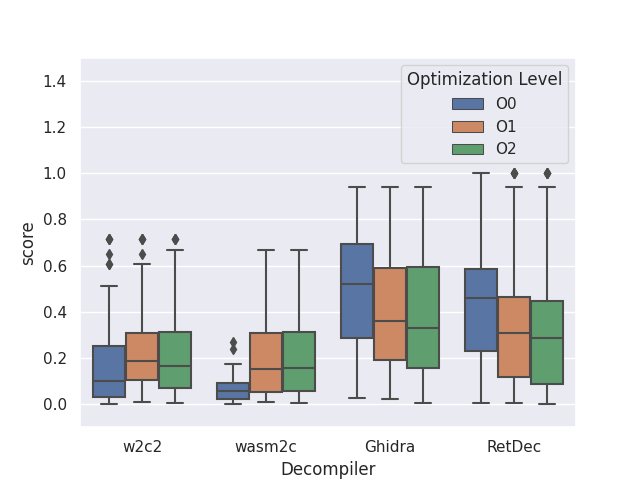} }}%
     \label{fig:plot_ted}
   \caption{Readability and structural similarity results consistent with Tables~\ref{tab:readability_results} and \ref{tab:similarity_results}.}
   \label{fig:results_plots}
\end{figure*}

%% file: Discussion.tex
\section{Discussion and Future Work}

\label{sec:discussion}

In this study, we evaluated and compared state-of-the-art WASM decompilers. 
To achieve our goal, we collected and integrated multiple metrics from previous works and created a comprehensive framework for evaluating decompilers. This framework assesses correctness, readability, and structural aspects, and includes a case study.
Through our analysis, we gained valuable insights into the performance and capabilities of existing decompilers. 
However, our investigation also revealed certain limitations and disadvantages that warrant attention in future research endeavors.

\subsection{Correctness}
We evaluated the correctness of WASM decompilers that generated compilable C code. Still, the process was limited by the absence of a substantial portion of the original features in C programming.
It would benefit the community if a framework were developed to assess the correctness of decompilers for a broader range of C programs.

Further investigations could also explore the correctness of WASM decompilers for different languages, platforms, or applications, extending the scope of our findings. 
Assessing our findings' generalizability across different WASM compilers and decompilers is crucial to comprehensively understanding correctness challenges.

\subsection{Readability and Structural Similarity}
We encountered several challenges when delving further into the readability and structural similarity. 
Distinguishing whether an issue lies within the scope of the ``WASM decompiler'' or is a broader ``WASM'' or ``decompiler'' problem posed a difficulty. 
For example, the inherent design of WASM as a stack-based machine lacks concepts of arrays. 
Consequently, a straightforward comparison between the decompiled code and native binaries becomes problematic due to their structural differences.
Devising fair evaluation methods to handle such disparities and identifying the appropriate properties for comparison is a compelling direction.

\subsection{Benchmark and Evaluation Metrics}
We conducted the experiments by treating the compile and decompile process as black-box, 
i.e., we naively compile C programs into WASM, decompile WASM binaries, execute them with evaluation metrics, and analyze results.
The benchmarks we chose are commonly used datasets for C programs, which cover different implementations of algorithms for a wide range of usages.
Though our approach is straightforward and unbiased, it may also strive away from the natural usage of WASM.
However, reproducing our approach with real-world WASM programs is difficult as most WASM programs are closed-source without ground truth for comparing the decompiled code against. 
One possible future direction may be a case-oriented study of the decompilation results for certain domain-specific usage of WASM, or analyzing the insufficiency of WASM decompilers from fundamental programming language aspects.

%% file: Related.tex
\section{Related Works}
\label{sec:related}

\noindent
\textbf{Reverse Engineering.}
Decompilation reverses the compilation and transforms the binary code back to high-level code. Since compilation is not a fully reversible process, it is difficult to construct a decompiler that can decompile an arbitrary executable back into the high-level code it was compiled from \cite{Miecznikowski_Hendren_2002}.
Reverse engineering can be used to find bugs, vulnerabilities, and malware, and perform verification, comparison, and multi-platform portability \cite{EelcoVisser_2023}.
Besides the two open-source native binary decompilers \textit{Ghidra}~\cite{ghidra} and \textit{RetDec}~\cite{retdec} in the study, we also considered angr \cite{Angr_2023} and snowman \cite{Yegord_2023}. However, they can only generate C pseudocode or C programs containing syntax errors; thus, decompiled programs are not recompilable. Thus, these native decompilers are excluded from the study.
To measure the complexity of decompiled programs, in addition to the widely-used Cyclomatic complexity metric~\cite{mccabe1976complexity}, previous decompiler research by Khaled et al.~\cite{yakdan2015no} highlighted that complicated goto statements in decompiled C code could be a significant obstacle for developers to understand the control flow. 
To improve readability, they attempted to reduce the number of such statements.
Subsequently, a later work by Eric et al.~\cite{schulte2018evolving} introduced more code properties, 
including the number of casts and dead assignments.

\noindent
\textbf{Reverse Engineering on WebAssembly.}
Wasmdec \cite{Wwwg_2023} is an open-source WebAssembly decompiler. However, it stopped updating in 2018 and didn't support many new features of WebAssembly standards. JEB \cite{Software_2023} is a commercial decompiler that provides the WebAssembly decompilation function. \cite{brandefelt2022decompilation} implemented a Datalog-based WebAssembly decompiler and found that all generated programs can be decompiled. More than 97\% of decompiled programs are recompilable, while only 70\% of the lowest complexity programs maintained correctness, and when the complexity increased, this percentage fell below 20\%. \cite{benali2022initial} investigated the viability of applying machine learning techniques, i.e., Neural Machine Translation (NMT), for decompiling WebAssembly binaries to C source code.

\noindent
\textbf{WebAssembly Study.} 
WebAssembly has been used for cryptomining \cite{konoth2018minesweeper, battagline2019hands, musch2019thieves}, games \cite{battagline2019hands}, software libraries \cite{narayan2019gobi, jeong2018watt}, computer vision \cite{opencvjs-doc, yuan-2020}, and encryption \cite{attrapadung2018efficient}. Researchers have analyzed topics including the presence of WebAssembly in the wild \cite{attrapadung2018efficient}, bugs in WebAssembly \cite{romano2021empirical}, and the performance of WebAssembly. 
To analyze WebAssembly performance, WebAssembly implementations of programs are compared with native or JavaScript counterparts. Haas et al.~\cite{haas_bringing_2017} compared the performance of WebAssembly to asm.js and native code. Jangda et al.~\cite{jangda2019not} investigated the slowdown of WebAssembly compared to native code and found that while optimizations can reduce the program size and execution time, some limitations arise from the design constraints of WebAssembly itself. Spies and Mock~\cite{spies_evaluation_2021} demonstrated that, on average, WebAssembly code is smaller and executes faster than equivalent JavaScript code. Yan et al.~\cite{yan2021understanding} compared the performance of WebAssembly with equivalent JavaScript and noted the effects of compiler optimization techniques, JIT optimization, and different platforms on the results. Compared with those WebAssembly topics, WebAssembly decompilers are less studied.

%% file: Conclusion.tex
\section{Conclusion}
\label{sec:conclusion}

In this paper, we performed an empirical analysis using 
a selection of diverse metrics to evaluate the generated code from WASM decompilers.
Our evaluation framework covers multiple perspectives, including correctness, readability, and structural aspects. 
By employing these metrics, we have demonstrated their usability in assessing C-based decompilers and provided valuable insights into the properties and limitations of current decompiled code.
We believe that our findings and the framework we presented will serve as a useful guide for future researchers in the field. 
Our work aims to foster the development of more sophisticated decompilers and contribute to the enhancement of both decompiling and WebAssembly toolchains.